\documentclass[12pt]{article}
\usepackage{enumitem}
\usepackage[english]{babel}
\usepackage[utf8]{inputenc}
\usepackage{algorithm,color}
\usepackage[intlimits]{amsmath} 
\usepackage{amsthm,mathrsfs}    
\usepackage[comma,authoryear]{natbib}
\usepackage{url}
\numberwithin{equation}{section}
\usepackage{amssymb,amsmath}
\usepackage{subfigure}
\usepackage{bm}
\usepackage{booktabs} 
\usepackage{graphicx,graphics,epsfig}

\usepackage{float}
\usepackage{multirow}
\usepackage{array}
\usepackage{setspace}

\newcommand{\Fb}{F_{\rm{B}}}
\newcommand{\fb}{f_{\rm{B}}}

\theoremstyle{plain}
\newtheorem{thm}{Theorem}

\newcommand{\bthm}{\begin{thm}}
\newcommand{\ethm}{\end{thm}}

\newcommand{\bpf}{\begin{proof}}
\newcommand{\epf}{\end{proof}}

\theoremstyle{definition}
\newtheorem{defn}{Definition}
\newtheorem{rem}{Remark}

\usepackage[margin=.88in,a4paper]{geometry}
\usepackage[compact]{titlesec}
\titlespacing*{\section}{0pt}{1.4pt}{1.4pt}
\titlespacing*{\subsection}{0pt}{1.4pt}{1.4pt}
\titlespacing*{\subsubsection}{0pt}{1.4pt}{1.4pt}
\renewcommand{\baselinestretch}{1.5}
\newcommand{\Lbeta}{\Leg_j \hspace{-.14em}\circ \hspace{.1em} F_B}
\usepackage{deep-macro}
\begin{document}
\begin{center}
{\Large{\bf Decentralized Nonparametric Multiple Testing}}\\[.15in] 
Subhadeep Mukhopadhyay\\
Department of Statistical Science,  Temple University\\ Philadelphia, Pennsylvania, 19122, U.S.A.\\[1.6em]
\end{center}
\begin{abstract}
Consider a big data multiple testing task, where, due to storage and computational bottlenecks, one is given a very large collection of p-values by splitting into manageable chunks and distributing over thousands of computer nodes. This paper is concerned with the following question: How can we find the \emph{full data multiple testing solution} by operating completely independently on individual machines in parallel, without \emph{any} data exchange between nodes?  This version of the problem tends naturally to arise in a wide range of data-intensive science and industry applications \emph{whose methodological solution has not appeared in the literature to date}; therefore, we feel it is necessary to undertake such analysis. Based on the nonparametric functional statistical viewpoint of large-scale inference,  started in \cite{deep16LSSD}, this paper furnishes a new computing model that brings unexpected simplicity to the design of the algorithm which might otherwise seem daunting using classical approach and notations.
\end{abstract} \vspace{-.4em}
\doublespacing
\setlength{\parskip}{1.2ex}
\section{The Open Problem} 
\begin{figure*}[t]
\centering
\includegraphics[height=.27\textheight,width=\textwidth,,keepaspectratio,trim=4cm 1.5cm 4cm 3cm]{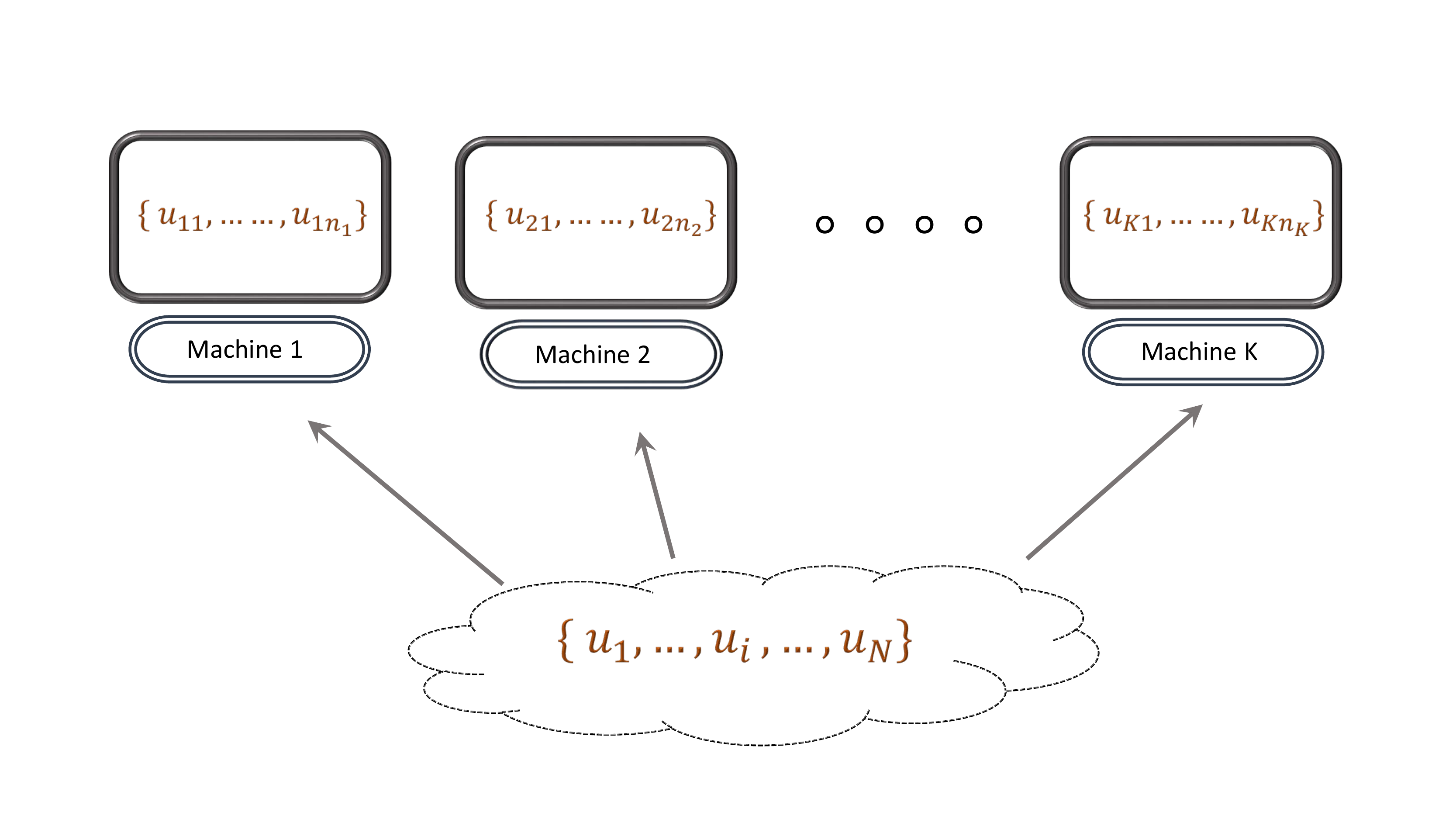}
\caption{The data structure and setting of decentralized large-scale inference problem. Massive collection of p-values distributed across large number of computer nodes.}
\label{fig:MTsetup}
\end{figure*}

Consider a multiple testing task with number of hypotheses in the millions, or even billions, as in high-throughput genomics, neuroscience, astronomy, marketing and other data-intensive applications. In this paper, we are interested in cases where these massive collection of p-values (corresponding to the null hypotheses) are distributed across multiple machines by breaking them into manageable chunks, as shown in Fig \ref{fig:MTsetup}. Given this set of partitioned p-values $\cP_j =\{u_{j1},\ldots,u_{jn_j}\},\,(j=1,\ldots,K)$, suppose the goal of a data scientist is to get the \emph{full data} (oracle) multiple testing result controlling overall false discovery rate (fdr), \emph{without} shipping all the p-values to a centralized computing machine, as this would clearly be unrealistic due to huge volume (too expensive to store), computational bottleneck\footnote[2]{BH \citep{BH95} and HC \citep{donoho2004} procedures start by ordering the p-values from smallest to largest incurring at least $O(N\log N)$ computational cost and other method like local fdr  \citep{efron01}  is of even greater complexity $O(N^2)$, thereby making legacy multiple testing algorithms infeasible for such massive scale inference problems.}, and possible privacy restrictions. Driven by practical need, the interest for designing \emph{Decentralized Large-Scale Inference Engine} has enormously increased in the last few years, due to their ability to scale cost effectively as the data volume continued to increase by leveraging modern distributed storage and computing environments.
There is, however, apparently \emph{no explicit algorithm currently available} in the literature to tackle this innocent-looking problem of breaking the multiple testing computation into many pieces, each of which can be processed completely independently on individual machines in parallel.
\begin{rem}
To get a glimpse of the challenge, consider a specific multiple testing method, say the Benjamini Hochberg's (BH) FDR controlling procedure, which starts by calculating the \emph{global}-rank of each p-value:
\vspace{-.45em}
\[\text{Global-rank of $u_{ji}$} = \text{\# p-values $\le u_{ji}$ in the \emph{full}-data} \cup_{j=1}^k \cP_j. 
\vspace{-.45em}\]
The computation of global-ranks, from the partitioned p-values, without \emph{any} communications between the machines, is a highly non-trivial problem. Difficulty with similar caliber also arises in implementing local false discovery type algorithms.
\end{rem}
The aim of this paper is to provide a general framework for designing Decentralized Large-Scale Inference algorithms, by adopting the nonparametric functional statistical viewpoint proposed in \cite{deep16LSSD}. The key to our theory is a new modeling principle, called the ``Superposition property,'' as a basis for addressing the big data challenge in a way that is easy to implement and understand (also teach).
\section{The Method}
In this paper, we suggest a new modeling theory for designing the desired scalable simultaneous inference architecture. At its core, there is a key representation scheme based on Superposition principle. To get there, however, we need to introduce a few modern notations and basic definitions.
\subsection{Background and Notations}
We begin by recalling the basic notations and some theoretical background as given in \cite{deep16LSSD}, which will be used throughout the paper. Let $Z_i$'s are the test statistic for the corresponding hypothesis testing problem $H_{i}$ ($i=1,\ldots,N$) and the goal is to detect false null hypotheses by testing them simultaneously. More broadly, we can think $Z_1,Z_2,\ldots,Z_N$ as a \emph{mixed} random sample, with the majority of the observations coming from null distribution $F_0$, and a small proportion from unknown signal distribution $H$: $F=\eta F_0 + (1-\eta) H,\, 0< \eta \le 1$. Note that here $H$ is arbitrary, i.e., it could be a mixture of unknown distributions of any complexity. 
\begin{defn}
Our nonparametric theory of large-scale inference starts by defining comparison distribution function between $F_0$ and $F$ (with respective densities $f_0$ and $f$) by $D(u;F_0,F):=F(F_0^{-1}(u))$ and the corresponding comparison density:
\[
d(u;F_0,F)\,=\, \dfrac{f(F_0^{-1}(u))}{f_0(F_0^{-1}(u))}, \quad 0 < u<1.
\]
\end{defn}
Consider testing $N$ independent null hypothesis $H_{1},\ldots,H_{N}$ based on corresponding p-values $u_1,\ldots,u_N$, where $u_i$ is equals to $F_0(z_i)$ or $1-F_0(z_i)$ depending on whether we want left-tailed or right-tailed p-values. If all the null-hypotheses are true (i.e., under $H_0:F=F_0$) we would expect $\widetilde D(u_i;F_0,F)= \widetilde F(F_0^{-1}(u_i)) \approx u_i$ where $\widetilde{F}(z;Z)=N^{-1}\sum_{i=1}^N \mathbb{I} (Z_i \le z)$. Thus, intuitively, one can suspect that the collection of p-values $\{u_i: \widetilde D(u_i)/u_i\,>\,\hat \ga\}$, for a suitably chosen threshold $\hat \ga$, data-dependent or constant, potentially correspond to the true signals or false null hypotheses. Based on this intuition, the following theorem presents an equivalent representation of the BH procedure \citep{BH95} in our notation:
\begin{thm}[Mukhopadhyay, 2016]
Let $u_{(1)} \le u_{(2)} \le \cdots u_{(N)}$ be the sorted p-values of $H_{(1)},\ldots,H_{(N)}$. Then the procedure that rejects $H_{(1)},\ldots, H_{(k)}$ where
\beq \label{eq:CDBH}
k~=~\argmax_{i} \Big\{ \dfrac{\widetilde{D}(u_{(i)})}{u_{(i)}}\,\ge\, \dfrac{\eta}{\al}\Big\}.
\eeq
controls FDR at the level $\al$, regardless of the distribution of the test statistic corresponds to false null hypothesis.
\end{thm}
Another popular method, Higher Criticism \citep{donoho2004}, also admits comparison distribution representation. Reject $H_{(i)}$ for $i=1,\ldots,k$ where
\[
k ~=~\Big\{ 1\le i \le \al_0 N:\, \argmax_{i} \sqrt{N} \dfrac{\widetilde D(u_{(i)}) - u_{(i)} }{\sqrt{u_{(i)} (1-u_{(i)})}}  \Big\}, ~~\al_0 \in (0,1).
\]
Furthermore, frequentist and Bayesian large-scale inference algorithms can be connected using the theory of reproducing kernel Hilbert space (RKHS) of the limiting Brownian bridge process of the comparison distribution. Efron's empirical Bayes local false discovery \citep{efron01} can alternatively be represented in terms of the p-values using our notation as
\beq \label{eq:locfdr}
\fdr(u)\,=\, \Pr({\rm null} \mid U=u)\,=\, \eta/d(u;F_0,F),~~0<u<1
\eeq
which leads to following  procedure: reject all $H_i$ if $\widetilde d(u_i;F_0, F)>\eta/2 \al$. \cite{efron07} showed that under certain condition on the alternatives, this method controls size, or Type I errors at the desired level $\al$.

Thus a harmonious unification between different cultures of multiple testing is possible by recasting it into a nonparametric comparison density function approximation problem, thereby allowing a more convenient and concise description of the existing techniques.
\subsection{Towards Decentralized Model}
The functional statistical reformulation discussed in the earlier section provides us with the first impetus towards decentralizing multiple testing computing. Nonetheless, to develop the explicit strategy (of estimating comparison density), we need more. Traditional off-the-shelf nonparametric density estimation algorithms (e.g., kernel density estimation technique) faces stiff modeling challenges; see Supplementary Appendix B1 for more discussion. To address this, we introduce a \emph{specialized} nonparametric model, called skew-Beta model, that is amenable to distributed computing. This is a critical in parallelizing the computation across a large number of machines, with \emph{no} loss of accuracy.
\begin{defn}
The Skew-Beta comparison density model is given by:
\beq \label{eq:sbeta}
d(u;F_0,F)\,=\,\fb(u;\,\gamma,\be) \Big\{ 1+\sum_j \LP[j;\Fb,D]\, T_j(u;\Fb)   \Big\}, \quad \text{for}~ 0<u<1,
\eeq
where beta density and distribution are denoted by $f_{\rm{B}}$ and $F_{\rm{B}}$, respectively; $T_j(u;\Fb)$ are called LP-polynomials--\emph{Legendre polynomials of rank-transformed random variables}, given by $\Leg_j(\Fb(u))$. LP polynomials constitute a complete basis in the Hilbert space $\cL^2(F_B)$, which dictates the optimality of the stochastic expansion \eqref{eq:sbeta}.
\end{defn}
\begin{thm}
The generalized LP-Fourier coefficients of the skew-Beta nonparametric model \eqref{eq:sbeta} admit the following representation:
\beq \label{eq:LPc-fd} \LP[j;\Fb,D]\,=\,\Ex[\Leg_j(\Fb(U)); D]\,=\,\int_0^1 \Leg_j(\Fb(u)) \dd D(u;F_0,F).\eeq
\end{thm}
This suggests that the unknown coefficients of the model \eqref{eq:sbeta} can be rapidly computed by taking the mean of the $\Leg_j$ score functions \emph{evaluated at the beta-transformed p-values}:
\[
\LP[j;\Fb,\widetilde D]\,=\, N^{-1}\sum_{i=1}^N \Leg_j\big[ F_{\rm{B}}(u_i;\gamma,\be) \big].
\]
\begin{rem}
The method described so far is applicable when one can access all the p-values $\{u_1,\ldots,u_N\}$ at once on a single computer. We call this framework \emph{Centralized Simultaneous Inference Model}. However, as we have argued this may not be a scalable and flexible setting in the ``big data'' era. Next, we address this limitation by developing a theory of computation that can operate in parallel on the p-values distributed across multiple computers to yield the oracle \emph{full data} multiple testing solution.
\end{rem}
Let $K$ denote the number of partitions or the number of CPUs, each containing $n_l$ p-values $(u_{l1},\ldots, u_{ln_l})$. The full data comparison distribution function can be expressed as
\[D(u;F_0,\wtF)=\wtF(Q(u;F_0))=N^{-1}\sum_{l=1}^k\sum_{i=1}^{n_l} \ind(u_{li} \le u) = \sum_{l=1}^k \pi_l D(u;F_0,\wtF_l),\]
where $\pi_l=n_l/N$, and $Q(u;F_0)$ is the quantile function for null $F_0$. Often we will be using a shorthand notation $\wtD_l$ for $D(u;F_0,\wtF_l)$ (by a slight abuse of notation) for compactness.
\vskip.4em
\begin{thm} \label{thm:LPc}
The full data LP-Fourier coefficients \eqref{eq:LPc-fd} admit the following notable distributed representation
\beq \label{eq:LP-dis} \LP[j;\Fb,\wtD]= N^{-1}\sum_{l=1}^K\sum_{i=1}^{n_l}\Leg_j(\Fb(u_{li})) = \sum_{l=1}^K \pi_l \LP[j;\Fb,\wtD_l], \eeq
where $\LP[j;\Fb,\wtD_l]=n_l^{-1}\sum_{i=1}^{n_l} \Leg_j(\Fb(u_{li}))$.
\end{thm}
\vskip.4em
\begin{rem}[Large-scale inference for big data via local modeling]
As a consequence of Theorem \ref{thm:LPc}, one can perform ``local modeling''-- modeling by computing the LP-coefficients $\LP[j;\Fb,\wtD_l]$ based on the local p-values $(u_{l1},\ldots, u_{ln_l})$ in parallel, to yield the full data ``global'' LP-coefficients. This allows us to scale multiple testing problems for very large datasets on a cluster of machines, leveraging big data processing platforms such as Apache Hadoop or Spark.
\end{rem}

Substituting \eqref{eq:LP-dis} into \eqref{eq:sbeta}, we have the following representation of the comparison density
\bea \label{eq:cd-dc}
d(u;F_0,\wtF) &=& \fb(u;\gamma,\be)\Big[ 1+\sum_{l=1}^K \pi_l \sum_{j=1}^m \LP[j;\Fb,\wtD_l]  \Leg_j(\Fb(u;\gamma,\be))   \Big] \nonumber \\
&=& \sum_{l=1}^K \pi_l \fb(u;\gamma,\be) \Big[ 1+ \sum_{j=1}^m  \LP[j;\Fb,\wtD_l]  \Leg_j(\Fb(u;\gamma,\be))   \Big].
\vspace{-1em}
\eea

\subsection{Superposition Principle}
Define the locally estimated comparison density for the $l$-th partition as
\beq  \label{eq:cd-local} d(u;F_0,\wtF_l)=\fb(u;\gamma,\be) \Big[ 1+ \sum_{j=1}^m \LP[j;\Fb,\wtD_l]  \Leg_j(\Fb(u;\gamma,\be))  \Big].\eeq
Combining the LP representations \eqref{eq:cd-dc} and \eqref{eq:cd-local}, we get the following remarkable decomposition formula.
\begin{thm}[The superposition principle] \label{thm:LSP} Under LP-expansion, the oracle full-data based comparison density estimate can be represented as the weighted sum of the locally estimated comparison densities:
\beq \label{eq:LSP} d(u;F_0,\wtF) = \sum_{l=1}^K \pi_l  d(u;F_0,\wtF_l), ~~0<u<1.  \eeq
\end{thm}

\begin{rem}
The modeling paradigm based on the principle of superposition suggests that we can estimate the global (full data) comparison density function by properly stitching together all the ``local snapshots'' $d(u;F_0,\wtF_l)$ in a completely parallelized manner. Furthermore, it doesn't matter how the p-values are partitioned, as the final aggregated result \eqref{eq:LSP} will always agree in the end. This idea of \emph{decomposition over distributed data-blocks} is interesting in its own right as a means of developing parallelizable algorithms for statistical modeling. 
\end{rem}

\begin{rem}[Signal heterogeneity index] The shape of the individual comparison density estimates $d(u;F_0,\wtF_l)$ in \eqref{eq:LSP} are highly informative in revealing how heterogeneous (signal-rich) the different p-value partitions are. In fact, the homogeneous data-distribution hypothesis $H_0: F_1=\cdots=F_K$, can equivalently be rephrased in terms of equality of component comparison densities $H_0': d_1=\cdots=d_K$. Consequently, if data partitions result in significantly different estimates of $d_l$, that would indicate the possibility of heterogeneity; thus, it is naturally tempting to come up with a rapidly computable measure of \emph{signal heterogeneity index}. For each partition define the \texttt{H-statistic}:
\beq \label{eq:impi} H_l\,\leftarrow\,\sum_{j=1}^m\big| \LP_l[j;U,\wtD_l] \big|^2~=~ \sum_{j=1}^m \Big| n_l^{-1}\sum_{i=1}^{n_l} \Leg_j(u_{li}) \Big|^2;~~~(l=1,\ldots,K).
\eeq
\end{rem}
The rationale comes from the following theorem.
\begin{thm} \label{thm:dep} For any arbitrary $G$ with support $[0,1]$, the skew-G LP represented comparison density is given by $d(u;G,F)=g(u)\{1+\sum_j \LP[j;G,D]T_j(u;G)\}$. Define the  for Chi-square divergence between $D$ and $G$ too be $\chi^2(D||G)=\int \big[\frac{d(u)}{g(u)}-1\big]^2 \dd G$. Then Chi-square divergence, which measures how close $d$ is from $g$, admits the following expression:
\beq \label{eq:H}
\sum_j\big|\LP[j;G,D]\big|^2\Ex_G[\Leg_j^2(U)]\,+\,\sum_{j \ne k} \LP[j;G,D] \LP[k;G,D]\,\Ex_G\big[\Leg_j(U)\Leg_k(U)\big].
\eeq
\end{thm}
Important point to note: To measure the departure from uniformity (the null p-value distribution) by selecting $G$ to be uniform distribution $U[0,1]$ in Theorem \ref{thm:dep}, the general expression \eqref{eq:H} drastically simplifies as Legendre polynomials are orthogonal with respect to the uniform measure, thereby boiling down to $\sum_j |\LP[j;U,D]|^2$, which can be readily computed using \eqref{eq:impi} for different partitions. The \texttt{H-statistic} values can be used to find high-priority (discovery-prone) partitions for careful investigations.
\begin{rem}[Data-driven weighted multiple testing]
At this point, an astute reader may be wondering whether we can also use $H_l$ as our data-driven weights to increase the power of the multiple testing procedures. Indeed, these group-specific heterogeneity indices can be used for constructing weights by properly normalizing them:
\[w_l=(\pi_l)^{-1}\frac{H_l}{\sum_l H_l} ~~\text{such that\, $\sum_{l=1}^K \pi_l w_l=1$}.~~~~~\]
The empirical detection power can be increased significantly in a heterogeneous case by running partition-specific scanning with \emph{different thresholds}. For example, The rejection region of \eqref{eq:CDBH} can be modified based on data-driven weights as $\mathcal{R}=\cup_{l=1}^K \mathcal{R}_l$ where
\beq \label{eq:DDbh}
\mathcal{R}_l=\text{Collection of p-values in the $l$-th partition} \le \max_{1 \le i \le n_l} \big\{u_{(li)}: \dfrac{\widetilde{D}(u_{(li)})}{u_{(li)}}\,>\,\dfrac{\pi_0}{ w_l \al}\Big\}
\eeq
This refined weighted version is expected to increase the power \citep{WWestfall2004,ignatiadis2016} of the proposed distributed multiple testing procedure compared to their unweighted counterparts. This demonstrates how heterogeneity can be leveraged for designing powerful large-scale distributed signal detection algorithms. Obviously, instead of data-driven nonparametric weights, domain scientists can also assign weights to each of the partitions using prior scientific knowledge, or they can use some kind of fusion of both data-driven and science-driven weights.
\end{rem}
\subsection{The Algorithm}
We outline the steps of  our algorithm derived from the theory and ideas described in the previous section.
\begin{center}
\emph{Algorithm: Decentralized Nonparametric Multiple Testing Engine}
\end{center}
\vspace{-.5em}
\medskip\hrule height .65pt
\vskip.6em
\texttt{Step 1.} We start with the collection of p-values $\{u_{l1},\ldots,u_{ln_l}\}_{l=1}^K$ distributed across $K$ machines; $N=\sum_{l=1}^K n_l$ denotes the total number of the p-values (which could be in billions and thus can exceed the capacity of a single machine).
\vskip.55em
\texttt{Step 2.} For $j=1,2$ compute $M_j=\sum_{l=1}^K\pi_l M_j[\wtD_l]$, where $M_j[\wtD_l]$ denotes the $j$-th order sample moment of the p-values present in the $l$-th partition $n_l^{-1}\sum_{i=1}^{n_l} u_{li}^j$, and $\pi_l=n_l/N$.
\vskip.55em
\texttt{Step 3.} Compute the method of moment estimators of the parameters of beta distribution
\[\hat \gamma=\dfrac{M_1(M_1-M_2)}{M_2-M_1^2};~~\hat \be=\dfrac{(1-M_1)(M_1-M_2)}{M_2-M_1^2}.\]

\texttt{Step 4.} For $l=1,\ldots,K$: At each partition separately compute
\vskip.15em
~~~~\,\texttt{Step 4a.} $\LP[j;\Fb,\wtD_l]=n_l^{-1}\sum_{i=1}^{n_l} \Leg_j(\Fb(u_{li};\hat \gamma,\hat \be))$;
\vskip.15em
~~~~\,\texttt{Step 4b.} $\LP[j;U,\wtD_l]=n_l^{-1}\sum_{i=1}^{n_l} \Leg_j(u_{li})$;
\vskip.15em
~~~~\,\texttt{Step 4c.} $H_l=\sum_{j=1}^m\big| \LP_l[j;U,\wtD_l]\big|^2$.
\vskip.55em
\texttt{Step 5.} Using Theorem 3, for $j=1,\ldots,m$ compute $\LP[j;\Fb,\wtD]= \sum_l^K \pi_l \LP[j;\Fb,\wtD_l]$.
\vskip.55em
\texttt{Step 6.} Return the estimated full data comparison density:
\[\widehat d(u;F_0,F)\,=\,\fb(u;\,\hat \gamma,\hat \be) \Big\{ 1+\sum_{j=1}^m \LP[j;\Fb,\wtD]\, \Leg_j(\Fb(u;\hat \gamma, \hat \be))   \Big\}, \quad \text{for}~ 0<u<1\]
where recall that $f_{\rm{B}}$ and $F_{\rm{B}}$ respectively denote beta density and distribution function. Estimate the smooth nonparametric model by selecting the `significantly large' LP-coefficients using the method proposed in \citet[Sec 3.3]{deep16LSSD}. At this point one can even estimate the proportion of true null hypothesis by applying the Minimum Deviance Algorithm of  \citet[Sec 3.4]{deep16LSSD} on $\widehat d(u;F_0,F)$.
\vskip.55em
\texttt{Step 7.} Implement \eqref{eq:CDBH}-\eqref{eq:locfdr}: they are upgraded nonparametrically \emph{smooth} versions of BH, HC, and local FDR. See Appendix B2 for more details. 
\vskip.55em
\texttt{Step 8.} For more insights, return heterogeneity indices $H_1,\ldots,H_K$. Partitions with higher \texttt{H}-index get prioritized. Display the chart consisting of pairs of points $(l,H_l)$; see Section 3 for more details.
\vskip.55em
\texttt{Step 9.} Further enhancement: Improve the power of the decentralized multiple testing procedure (Step 7) by using partition-specific thresholds. Compute data-driven weights $w_l=(\pi_l)^{-1}\frac{H_l}{\sum_l H_l}$ $(l=1,\ldots,K)$ and incorporate into \eqref{eq:DDbh}.
\vskip1em
\medskip\hrule height .65pt
\vskip1.5em
\vskip.25em
\begin{rem} The proposed algorithm immediately allows parallelization and a MapReduce type implementation. In particular, the `\texttt{Map}()' function consists of Steps 2 and 4 (local modeling and parallel execution); and in the `\texttt{Reduce}()' stage we perform (combining) Steps 3,5, and 9 (requires \emph{no} data exchange between nodes). As a result, our modeling framework represents a significant step forward, for it enables massive scalability to perform simultaneous inference on genuinely large datasets distributed over a cluster of commodity machines.
\end{rem}
\vspace{-.4em}
\section{Examples}
Two examples will be discussed one real data and the other one a simulated study.

{\bf Example 1}. Prostate cancer data \citep{prostate} consists of $102$ patient samples ($50$ normal and $52$ as prostate tumor samples) and $N=6033$ gene expression measurements. We aim to detect interesting genes that are differentially expressed in the two samples. For this purpose, we compute p-values based on two-sample t-test for each gene. Instead of having all the p-values in one centralized machine, we assume that they are distributed across $K$ processors based on the following partitioning scheme:  sort the p-values and separate the lowest 1\% of p-values ($\sim 60$ p-values) and randomly divide them into three blocks of equal size $\mathcal{L}_1, \mathcal{L}_2$ and $\mathcal{L}_3$.
\begin{figure*}[!htt]
 \centering
 \vskip.4em
 \includegraphics[height=.36\textheight,width=.46\textwidth,trim=4cm 0cm 4cm 0cm]{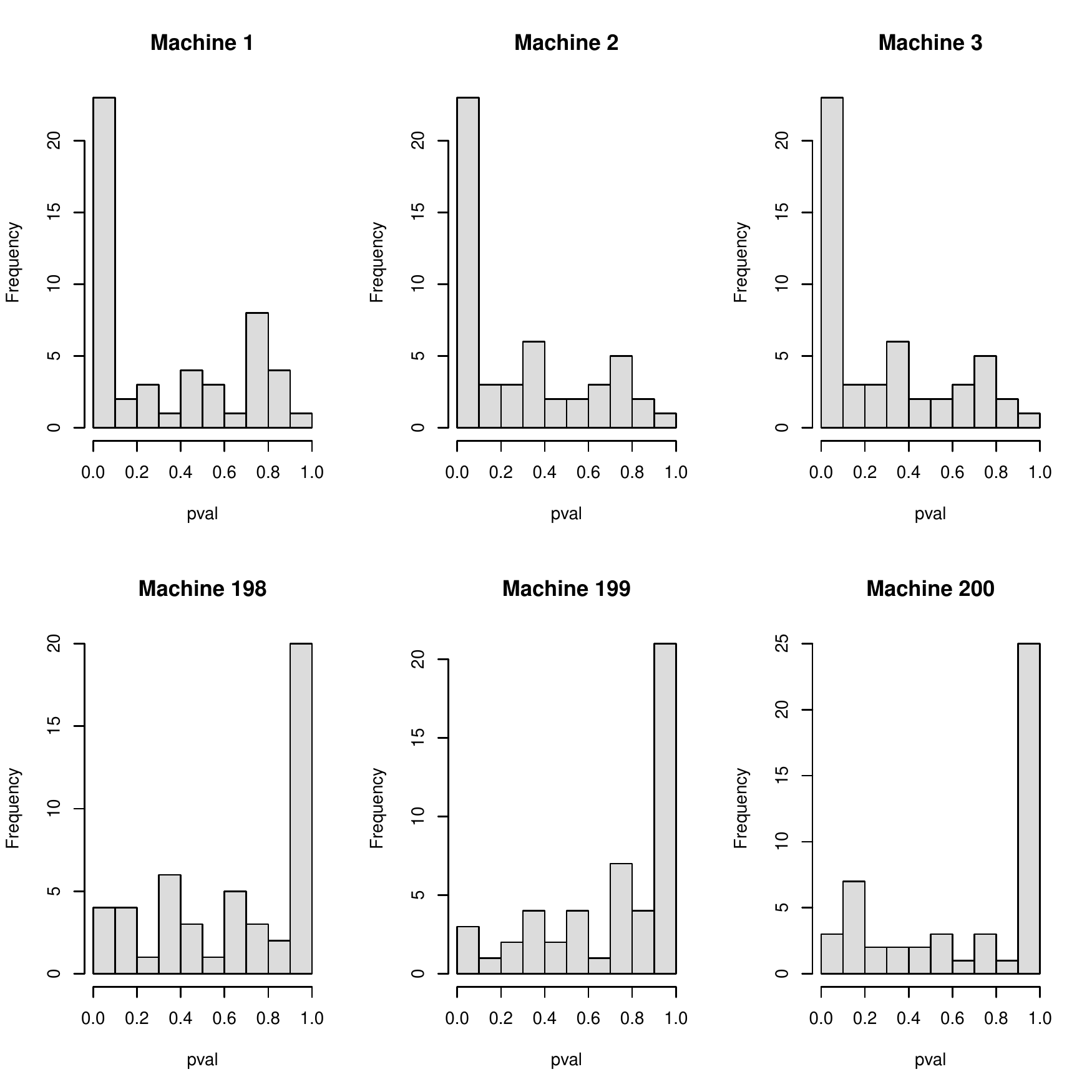}\\
 \includegraphics[height=.34\textheight,width=.64\textwidth,trim=0cm 0cm 0cm 1cm, clip=true]{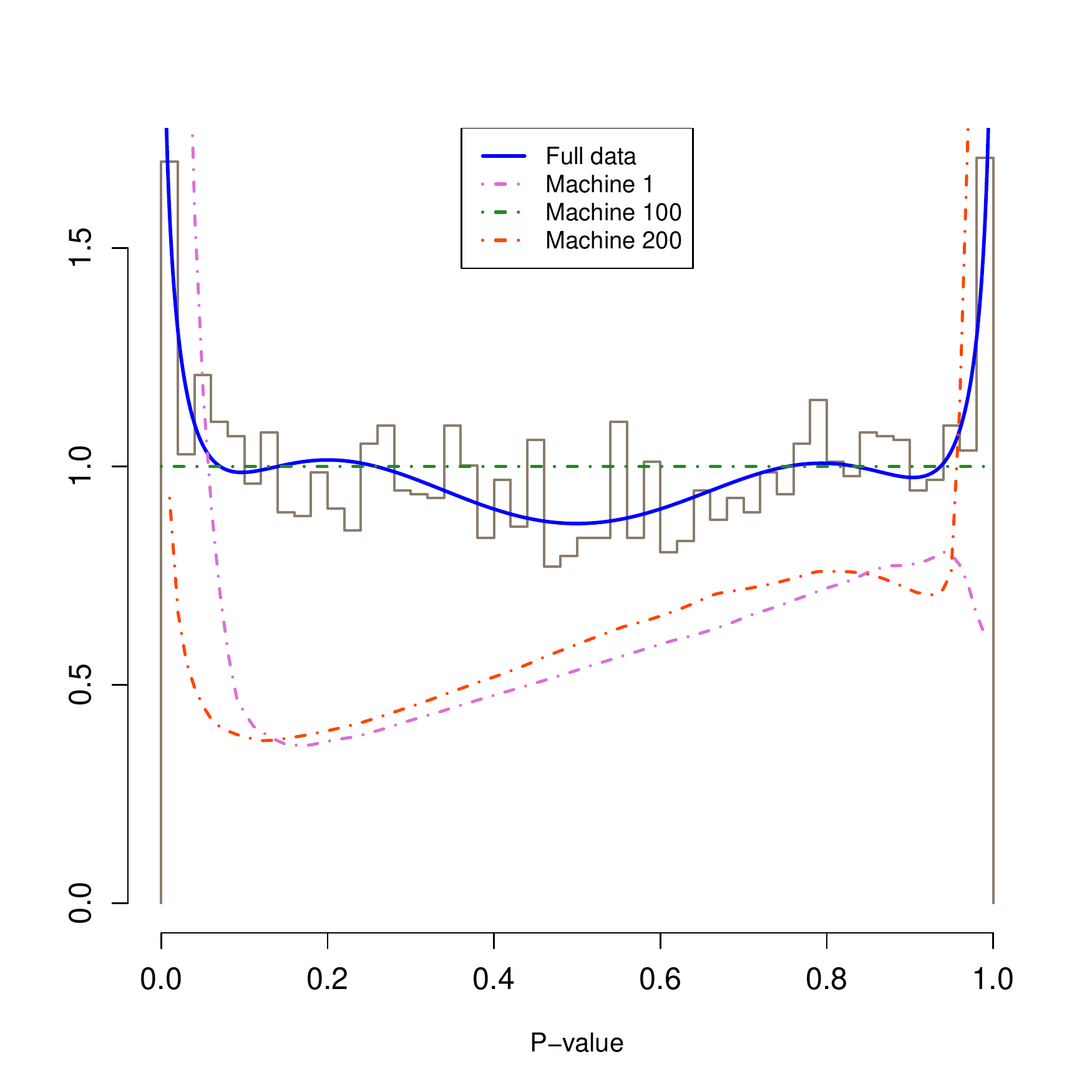}
\caption{(color online) Distribution of first and last three partitioned p-values for prostate data. Last row shows the $\widehat d_l$ $(l=1,100,200)$ along with the full-data $\widehat d$, computed using the superposition rule $\widehat{d}= \sum_{l=1}^K \pi_l  \widehat{d}_l$.} \label{fig:pros}
\end{figure*}
Do the same for the top 1\% of the p-values and create $\mathcal{U}_i$ $(i=1,2,3)$--each containing $20$ p-values. Randomly shuffle the rest of the p-values and bin them equally into $K=200$ partitions $\cP_1,\ldots,\cP_{200}$; and finally construct
\vspace{-.4em}
\[~~\cP_1 \cup \mathcal{L}_1,\, \cP_2 \cup \mathcal{L}_2,\, \cP_3 \cup \mathcal{L}_3,\,\, \cP_4, \ldots, \cP_{197},\,\, \cP_{198} \cup \mathcal{U}_1,\, \cP_{199} \cup \mathcal{U}_2,\, \cP_{200} \cup \mathcal{U}_3.~~~\]
Fig \ref{fig:pros} shows the distribution of p-values in the first and last three machines, which represents the  active components in \eqref{eq:LSP}. By design, all the remaining $194$ partitions have uniformly distributed p-values i.e. $d_l\equiv 1$. In what follows, we present a three-tier analysis pipeline: 
\vskip.65em
\texttt{Level 1}. We compute the LP-Fourier coefficients $\LP[j;\Fb,\wtD_l]$ in parallel mode at each of the computing nodes of the cluster. By combining all of them using Step 5, our algorithm yields the following \emph{full-data} comparison density estimate for the distributed prostate data:
\[
\dhat(u;\Phi,F)\,=\, .75 \,\big[ 1\,+\, 0.0589 \Leg_6 \big( F_{\rm{B}}(u;\widehat \gamma=.861,\widehat \be=.862) \big)  \big]\,u^{-.138}\,(1-u)^{-.137},~ ~~0<u<1.
\]
Although it is self-evident, it is important to point out that, irrespective of the partitioning scheme,  our algorithm is guaranteed to reproduce the same full-data result. We can now use this estimate at each partition to identify the discoveries by using \eqref{eq:CDBH}-\eqref{eq:locfdr} at the desired fdr level (also see Appendix B2).  For example, straightforward computation by applying \eqref{eq:locfdr} at $\al=.2$ finds $65$ non-null genes ($32$ in the left tail and $33$ in the right), spread over first and last three partitions. Whereas using two-sided p-values, $u_i=2 \Phi(-|z_i|)$ the \emph{smooth}-BH procedure by plug-in $\widehat D(u_{li})=\int_0^{u_{li}} \dhat(v;\Phi,F)\dd v $ in \eqref{eq:CDBH} declares $63$ genes ($30$ in the left tail and $33$ in the right) to be significant.
\vskip.65em

\begin{figure*}[t]
 \centering
 \includegraphics[height=.15\textheight,width=.32\textwidth,trim=2cm .5cm 1cm 2cm]{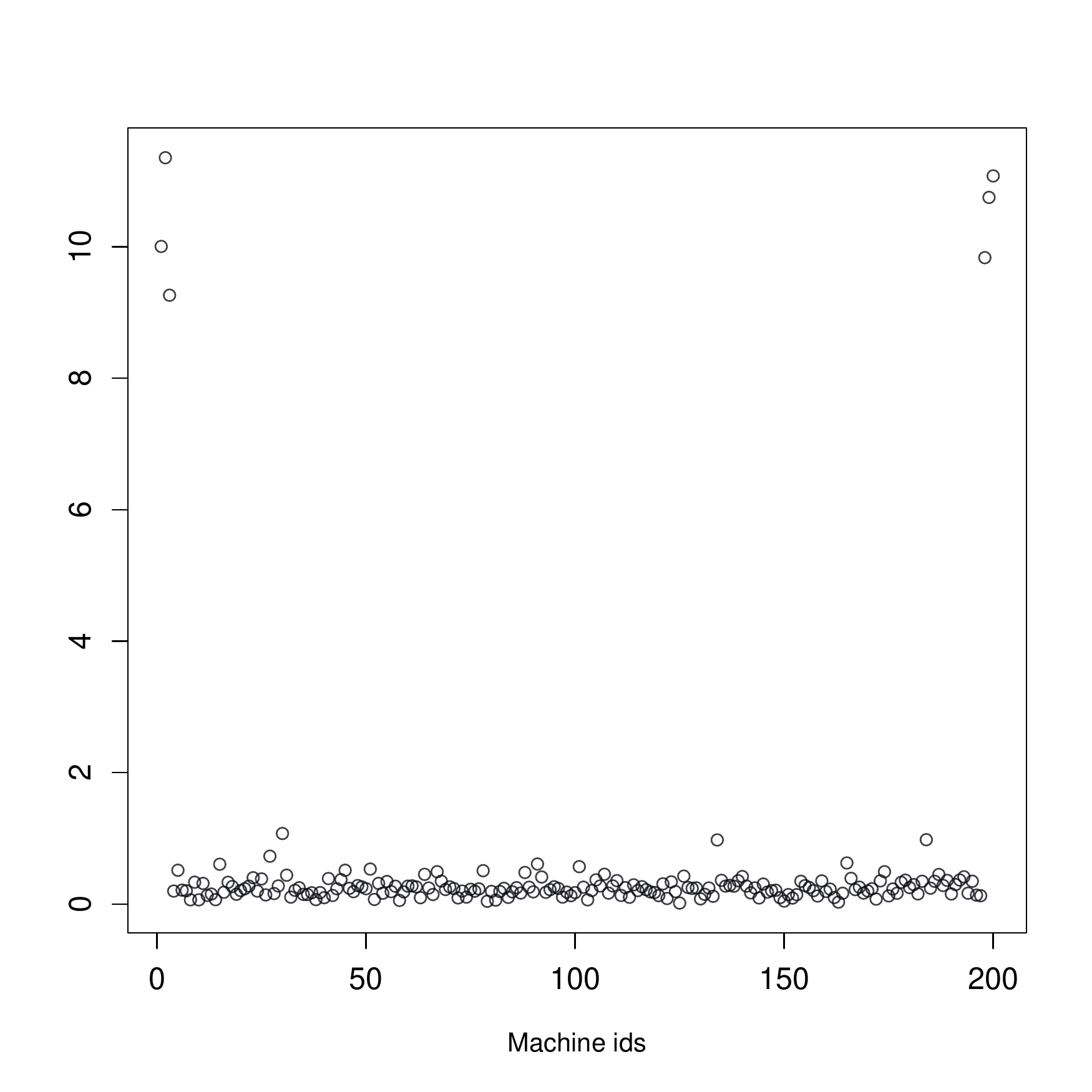}~~~~~~~~
 \includegraphics[height=.15\textheight,width=.32\textwidth,trim=1cm .5cm 2cm 2cm]{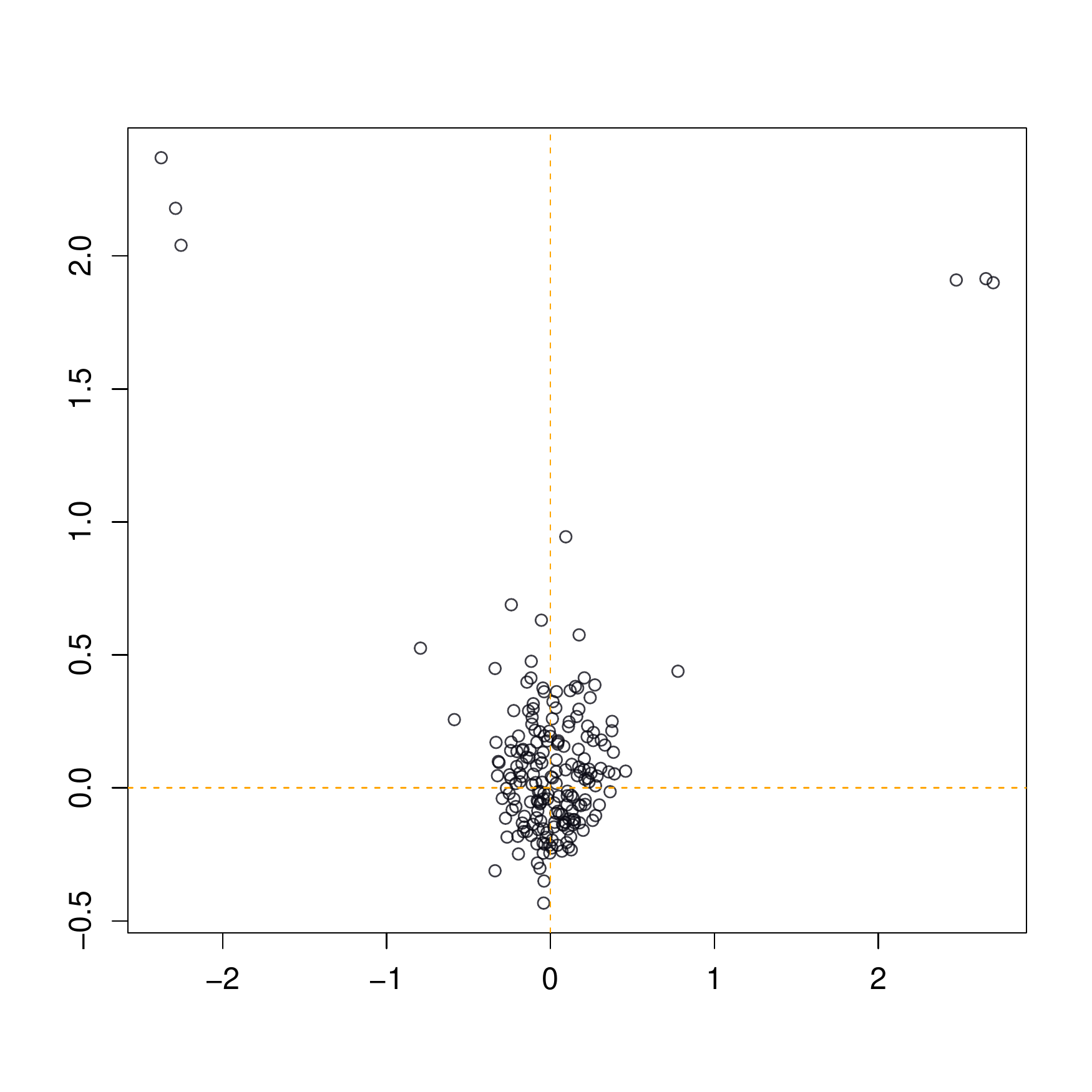}\\[.2em] \includegraphics[height=.15\textheight,width=.32\textwidth,trim=2cm 1cm 1cm 1.5cm]{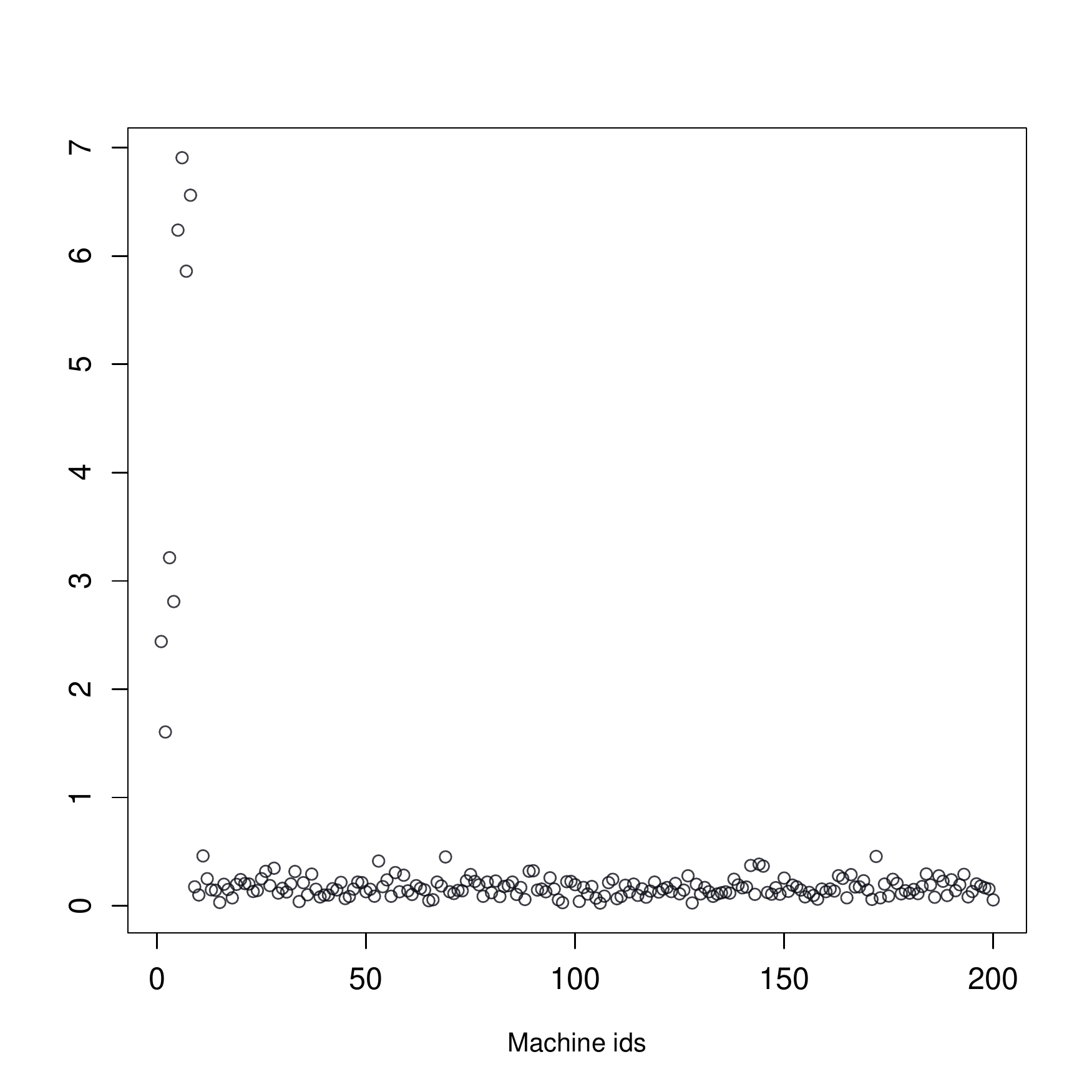}~~~~~~~~
 \includegraphics[height=.15\textheight,width=.32\textwidth,trim=1cm 1cm 2cm 1.5cm]{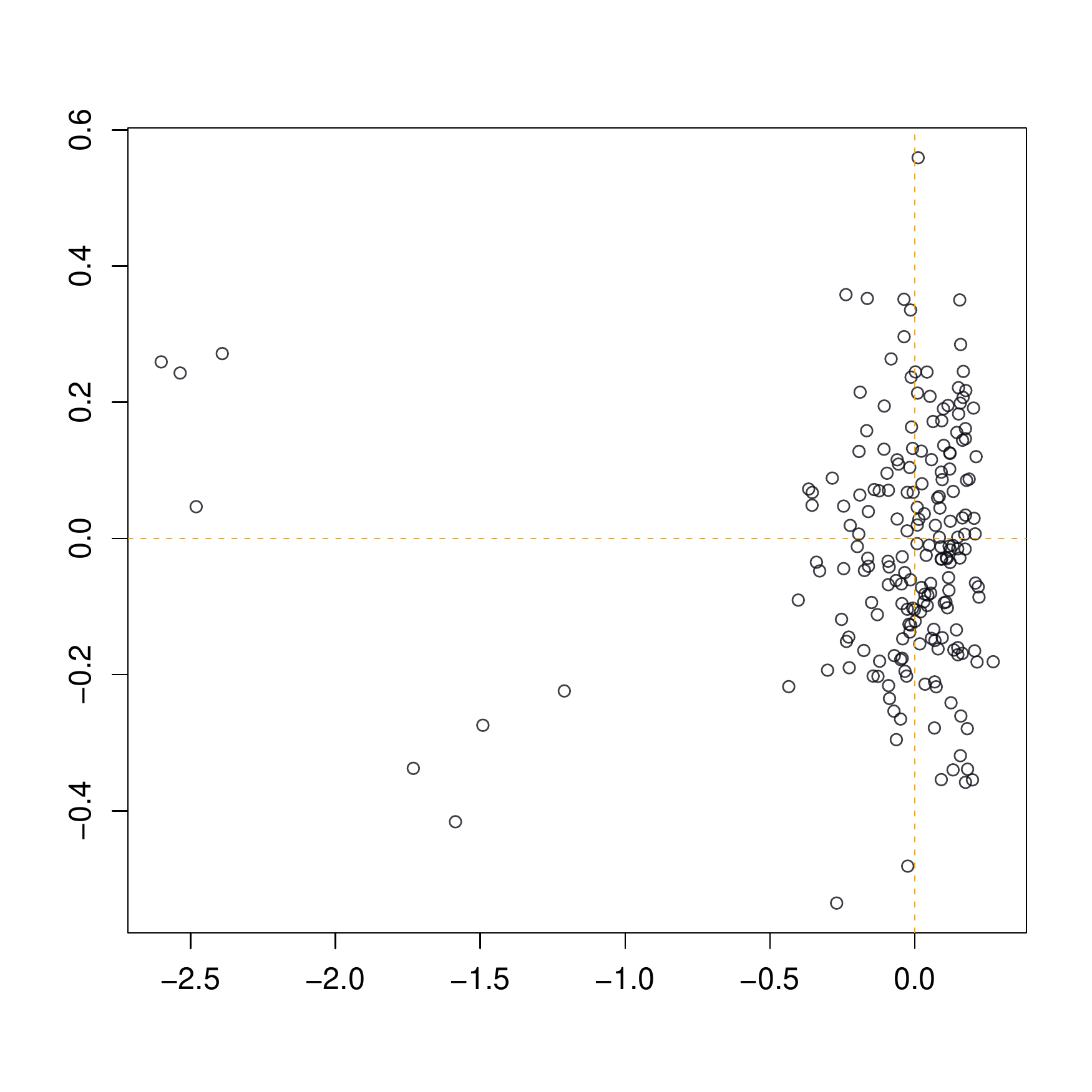}
\caption{Left panel: The Control H-chart (to identify signal-rich data sources); Right panel: The information Map (distance from origin signifies the importance of that partition). The first row displays the result for Prostate data and second row for Example 2.} \label{fig:pros2}
\end{figure*}

\texttt{Level 2}. The goal here is to identify the signal-rich p-value sources using \texttt{H}-index. Recall the first and last three partitions contain the discoveries, and are thus expected to have large H-statistic \eqref{eq:impi} value. The top left panel of Fig \ref{fig:pros2} plots the pair of points $(l,H_l)$, which we call ``Control H-chart.''  Use this chart to monitor and quickly spot the `informative' batch of p-values. For partitioned prostate data, as expected, the H-chart indicates that the first and last 3 groups are the primary source of discoveries (rejected null hypotheses).

Note that two different partitions may have similar value of H-indices, while the statistical characteristics might be very different. For example, in the prostate data the partitions $\{\mathcal{U}_j\}_j$ contain genes with large positive t-statistic (upper-tail); in contrast, the partitions $\{\mathcal{L}_j\}_j$ contain smallest (negative) t-statistic (lower-tail). Yet, as shown in Fig 3 (top left panel), the magnitudes of the H-statistic for both of the groups are comparable, in fact almost equal! Naturally at this point, an investigator may be interested in more refined grouping of the p-value sources \emph{based on signal characteristics}. To illustrate this point we introduce our second example.

\vskip.5em
{\bf Example 2}. Generate $9800$ samples from $\cN(0,1)$ and divide them equally across $K=200$ pieces. In each of the first four partitions we add $25$ samples from $\cN(2,1)$, and in the next four partitions we add samples generated from $U(2,4)$. Thus, among $200$ partitions, only the first eight contain the discoveries, albeit of two kinds. The H-chart shown the bottom left corner of Fig \ref{fig:pros2} correctly separates the  eight informative p-value sources from the rest.
\vskip.65em
\texttt{Level 3}.  Define the $k \times m$ L-matrix, with $L_{ij}=\LP[j;U,\wtD_i]$. Perform the singular value decomposition (SVD) of $L=U\Lambda U^{T}$ $= \sum_l \la_l u_l u_l^{T}$, where $u_{ij}$ are the elements of the singular vector matrix $U=(u_1,\ldots,u_m)$, and $\Lambda={\rm diag}(\la_1,\ldots,\la_m)$, $\la_1 \ge$ $ \cdots \la_m \ge 0$. Define the principal signal-profile coordinate of $i$th partition as by $\la_j u_{ij}$ for $j=1,\ldots,m$. The two-dimensional exploratory graph in the right panel of Fig \ref{fig:pros2} is formed using the points  $(\la_1 u_{i1}, \la_2 u_{i2})$, for $i=1,\ldots,k$ by taking the dominant two terms of the SVD. Different partitions or groups are displayed as points that allow us to separate the groups according to the statistical nature of signals.
\begin{rem}(Efficiency and accuracy dilemma). (i) the whole analysis can be done without moving any p-values across the data silos -- a cost-effective and accelerated computation. (ii) The proposed decentralized technique is an \emph{exact} method. The Superposition principle along with theorems 3-5 should be interpreted as identities that hold for \emph{any} arbitrary partitions (\emph{partition-invariance}), i.e. irrespective of how you break $N$ hypotheses into $K$ parts!
\vspace{-.4em}
\end{rem}
\section{Conclusions} 
Without losing the organic character of the general theory of nonparametric multiple testing proposed in \cite{deep16LSSD}, we successfully derived its non-trivial extension that allows transition from centralized to decentralized capability to scale for massive datasets with billions of tests. This shift is necessary in order to fully realize the potential for ever-increasing amounts of distributed big datasets, which has become the de facto standard in science, industry, and business. The core principles and ideas presented in this paper provide a comprehensive framework by embracing small (centralized) and massive (distributed) scale multiple testing cultures in a way that is intuitive and easy-to-implement; as a result, they have the potential to radically simplify theory, practice, and teaching. Prostate cancer data and simulated examples are used to illustrate the main steps (and more importantly the interpretations) of our algorithm. Obviously more complicated and large datasets could be used, but this should suffice to get the point across.
\section*{Acknowledgement}
The author would like to thank two anonymous reviewers for their helpful and constructive comments that greatly contributed to improving the final version of the paper. 
\section*{Disclosure statement}
No potential conflict of interest was reported by the author.
\section*{Supplementary Material}
Available online. Includes proofs of the main results and some additional details.
\bibliographystyle{asa}
\bibliography{ref-bib}
\newpage
\renewcommand{\baselinestretch}{1.34}
\setlength{\parskip}{.5ex}
\begin{center}
{\Large {\bf Online Supplementary Appendix for \\``Decentralized Nonparametric Multiple Testing''}}\\[.1in] %
Subhadeep Mukhopadhyay$^*$\\  
Temple University, Department of Statistical Science \\ Philadelphia, Pennsylvania, 19122, U.S.A. \\[.5em]
$^*$ Email correspondence should be directed to deep@temple.edu\\[2em]
\end{center}
This supplementary document contains two Appendices. Appendix A provides several proofs of results in the main paper. Appendix B includes some additional remarks.
\begin{center} 
{\large A. PROOFS}
\vspace{-.64em}
\end{center} 
\vskip.5em
{\bf A1. Proof of Theorem 2} 
\vskip.25em 
We start by noting the skew-beta model density model:
\beq \label{eq:sbetaapp}
d(u;F_0,F)\,=\,\fb(u;\,\gamma,\be) \Big\{ 1+\sum_j \LP[j;\Fb,D]\, T_j(u;\Fb)   \Big\}, \quad \text{for}~ 0<u<1, 
\eeq
where beta density and cdf with parameters $\gamma$ and $\be$ are denoted by $f_{\rm{B}}$ and $F_{\rm{B}}$, respectively; $T_j(u;\Fb)$ are called beta-LP polynomials $\Lbeta(u;\gamma,\be)$. Here the sign `$\circ$' refers to the usual composition of functions. The beta-LP polynomials satisfy the following orthonormality conditions:
\[\Ex_{F_B}[T_j(U;F_B)]=0,~~\text{and}~~\Ex_{F_B}[T_j(U;F_B)T_k(U;F_B)]=\delta_{jk}.\]
This implies that the LP-Fourier coefficients of \eqref{eq:sbetaapp} can now be expressed as
\bea \label{eq:sup:lpcoef}
\LP[j;\Fb,D]&=&\int_0^1 \dfrac{d(u;F_0,F)}{\fb(u;\,\gamma,\be)} T_j(u;F_B) \dd F_B(u;\gamma,\be) \nonumber \\ 
&=& \int_0^1  T_j(u;F_B)  \dd D(u;F_0,F) = \Ex_D[T_j(u;F_B)].
\eea  
Complete the proof by replacing the population $D$ in \eqref{eq:sup:lpcoef} by its sample estimator $\widetilde{D}$ to compute $\LP[j;\Fb,\widetilde{D}]$. \qed
\vskip.5em
{\bf A2. Proof of Theorem 3} 
\vskip.25em
We begin by recalling the definition of sample comparison density $\wtD_l \equiv D(u;F_0,\wtF_l)$ of the $l$-th partitioned p-values:
\beq \label{eq:Dl}
D(u;F_0,\wtF_l) \,=\, \wtF_l(Q(u;F_0))\,=\,n_l^{-1}\sum_{i=1}^{n_l} \ind(u_{li} \le u). 
\eeq 
Theorem 2 implies that the sample LP-Fourier coefficients for the $l$-th partition is given by
\beq \label{eq:LPl}
\LP[j;\Fb,\wtD_l] = n_l^{-1} \sum_{i=1}^{n_l}\Lbeta(u_{li};\gamma,\be), ~~j=1,\ldots,m.
\eeq 
This ensures that the full-data sample LP-Fourier coefficients can be expressed as 
\beas 
\LP[j;\Fb,\wtD]&=& N^{-1}\sum_{l=1}^K\sum_{i=1}^{n_l}\Lbeta(u_{li};\gamma,\be) \nonumber \\
&=& \sum_{l=1}^K \Big\{ N^{-1} \sum_{i=1}^{n_l}\Lbeta(u_{li};\gamma,\be)\Big\},
\eeas
which by virtue of \eqref{eq:Dl} and \eqref{eq:LPl}, can be rewritten as 
\[\LP[j;\Fb,\wtD] = \sum_{l=1}^K \pi_l \LP[j;\Fb,\wtD_l],\]
where $\pi_l=n_l/N$. This proves the claim. \qed
\vskip.65em
{\bf A3. Proof of Theorem 4} 
\vskip.25em
This is immediate from \eqref{eq:cd-local} and Theorem 3, as noted in \eqref{eq:cd-dc}.
\vskip.65em
{\bf A4. Proof of Theorem 5} 
\vskip.25em
The chisquare divergence between skew-G comparison density $$d(u;G,F)=g(u)\big\{1+\sum_j \LP[j;G,D]T_j(u;G)\big\},$$ and an arbitrary $G$ over the unit interval is given by 
\beq \label{eq:CHI}
\chi^2(D||G)=\int_0^1 \left[\frac{d(u)}{g(u)}-1\right]^2 g(u) \dd u 
=\int_0^1 \left\{  \sum_j \LP[j;G,D]T_j(u;G) \right\}^2 g(u) \dd u.
\eeq
Straightforward calculation shows \eqref{eq:CHI} has the following analytic form:
\[\sum_j\big|\LP[j;G,D]\big|^2\int_0^1 T_j^2(u;G) \dd G\,+\,\sum_{j \ne k} \LP[j;G,D] \LP[k;G,D]\,\int_0^1 T_j(u;G) T_k(u;G) \dd G,\]
which completes the proof. \qed
\vskip.65em
\begin{center} 
{\large B. ADDITIONAL REMARKS}
\vspace{-.25em}
\end{center} 
\vskip.25em
{\bf B1. Advantages of LP-skew Density Model}. The reason for using LP-skew density model \eqref{eq:sbeta} instead of classical kernel density estimate (KDE) is threefold: 
\begin{itemize}[itemsep=2pt,topsep=2pt]
\item \textit{Statistical side}: KDE for compact support $[0, 1]$ is known to be
a challenging problem due to the ``boundary effect,'' Besides this, difficulty arises to accurately estimate the \textit{highly dynamic} tails near $0$ and $1$, such as shown in the bottom panel of Fig 2. As noted in \cite{deep16LSSD}, the novelty of our approach lies in its unique ability to “decouple” the density estimation problem into two separate modeling problems: the tail part and the central part of the distribution. Keep in mind that tails (where the signals hide)  of $\widehat{d}(u;F_0,F)$ are the most important part for multiple testing.

\item \textit{Computational side}: The brute-force application of KDE $\frac{1}{Nh}\sum_{l=1}^K\sum_{i=1}^{n_l} K\left( \frac{u-u_{li}}{h}\right)$ requires $O(N^2)$ kernel evaluations and $O(N^2)$ multiplications and additions, making it computationally impractical for large-$N$ problems (even for a fixed-bandwidth case).
\item \textit{Compressibility side}: The skew-beta model encodes the shape of the density using few LP-Fourier coefficients\footnote[2]{Note that, our specially designed LP-basis functions $T_j(u;F_B)$ are: (i) orthonormal basis with respect to the measure $F_B$, which guarantees parsimony of our density expansion, and (ii) robust in nature (as they are polynomials of rank-transform $F_B(u;\gamma,\be)$, thus can tackle highly-dynamic tails of the distribution without falling prey to the spurious bumps.}. For example, in the Prostate cancer example, we were able to compress the whole function into three coefficients. This compressive representation is particularly attractive for designing memory-efficient big-data algorithms. Contrast this with classical KDE approach, where storing the density estimate values at each data point could be expensive, if not infeasible.  
\end{itemize}
\vskip.5em
{\bf B2. On The Algorithm}. The prescribed embarrassingly parallel inference algorithm: 
\begin{itemize}[itemsep=2pt,topsep=2pt]
\item Upgrades traditional raw-empirical multiple testing procedures to a more stable and smooth-nonparametric versions.
\item Performs smooth-BH filtering, by computing $u_{{\rm max}}=\sup_u \Big\{  \frac{\widehat{D}(u)}{u} \ge \dfrac{\eta}{\al}\Big\},$
which can be done without any reference to the partitioned-pvalues once we have the $\widehat{D}$. Report the cases with $u_{li} \le u_{{\rm max}}$ as interesting for $l=1,\ldots,k$. Contrast this with the ``naive'' $\widetilde{D}(u)$ based BH procedure \eqref{eq:CDBH}, which requires sorting of p-values to count the empirical proportions. Also see Remark 1. 
\item Along the same line, one can also perform local-fdr analysis by evaluating $\widehat d(u_{li};F_0, F)>\eta/2 \al$ inside each partition, once we have $\widehat d$ (computed in a completely parallelized manner with zero-communication between the  nodes).
\end{itemize}
This again shows the usefulness of comparison-density-based functional reformulation of multiple testing problems.
\vskip.5em
{\bf B3. Functional View of Multiple Testing}. As noted in \cite{deep16LSSD}, the notion of comparison distribution allows us to transform the simultaneous hypothesis testing problem into a nonparametric function estimation problem. The transition from discrete analysis and ranking of individual p-values to comparison density function estimation\footnote[2]{This can also be viewed as going from large-$N$ microscopic discrete model to a functional macroscopic model that obeys the superposition principle (see Remark 4 of the main paper).} is necessary to develop the decentralized large-scale inference (DSLI) engine.

\vskip.8em
{\bf B4. Model Selection}. For constructing skew-beta model it is important to properly select the empirical LP-Fourier coefficients appearing in \eqref{eq:sbetaapp}. Identify indices $j$ for which $\LP(j; F_B; D)$ are significantly non-zero by using AIC model selection criterion applied to LP means arranged in decreasing magnitude. Choose $k$ to maximize ${\rm AIC}(k)$,
\[ {\rm AIC}(k)\,=\,\text{sum of squares of first $k$ sorted LP-means}\,-\,2k/N\]
This functionality was incorporated as an inbuilt mechanism for our decentralized algorithm. From a theoretical perspective, the proposed AIC-based LP-Fourier coefficient selection criterion can be shown to minimize the mean integrated squared error \citep[Sec. 2.4]{Deep17LPMode}.

\vskip.5em
{\bf B5. Real Examples of Massive-Scale Inference}. Modern data-intensive sciences and engineering applications routinely generate huge-scale inferences.  
\vskip.25em
The following are the two examples from genetics, where millions to billions of hypotheses are tested routinely to perform multiple hypotheses testing procedures. The first one is GWAS (or even microbiome-wide association) studies \citep{gwas,grubert2015genetic}, which require procedures that can perform tens of billions of tests for finding significant interaction between the pairs of single-nucleotide polymorphisms (SNPs) within a reasonable timeframe. The second example is eQTL studies \citep{eqtl}, usually consist of $10^9$ tests. No doubt there are innumerable examples like this, which necessitate a distributed multiple testing architecture.  

\vskip.5em
{\bf B6. Same Covariates on Different Machines}. Consider the case where we have same covariates on different machines. Define $\bar{x}_0$ and $\bar{x}_1$ to be the global group-specific sample means, which can be computed easily (in a parallelized manner):
\[\bar{x}_0 = \sum_{l=1}^k \pi_{l0}\bar{x}_{l0}, ~\text{and}~\bar{x}_1 =  \sum_{l=1}^k  \pi_{l1}\bar{x}_{l1},\]
where $\pi_{l0}=n_{l0}/N_0$, $\pi_{l1}=n_{l1}/N_1$, $n_l = n_{l0} + n_{l1}$, $N_0= \sum_{l=1}^k  n_{l0}$, and $N_1= \sum_{l=1}^k  n_{l1}$. Exact similar process is also valid for the sample standard deviations $S_1^2$ and $S_2^2$. This implies that we can easily compute the full-data Z or t-statistics $Z_1,\ldots, Z_p$ and can perform multiple-testing without any problem. 
\vskip.25em
On the other hand, this paper addresses the challenging regime where a massive collection of covariates are distributed over the machines, which needs a non-trivial solution and carries more appeal than the `large-n small-p' case, especially in the context of multiple testing.

\end{document}